\documentclass[%
 aip,
 chaos,
 amsmath,amssymb,
 reprint,%
]{revtex4-1}

\usepackage{graphicx}% Include figure files
\usepackage{dcolumn}% Align table columns on decimal point
\usepackage{bm}% bold math
\usepackage[utf8]{inputenc}
\usepackage[T1]{fontenc}
\usepackage{etoolbox}
\usepackage{xcolor}
\usepackage{hyperref}

\makeatletter
\def\@email#1#2{%
 \endgroup
 \patchcmd{\titleblock@produce}
  {\frontmatter@RRAPformat}
  {\frontmatter@RRAPformat{\produce@RRAP{*#1\href{mailto:#2}{#2}}}\frontmatter@RRAPformat}
  {}{}
}%
\makeatother
\begin{document}

\preprint{AIP/123-QED}

\title{Inertia-induced mechanism for giant enhancement of transport generated by active fluctuations}
% Force line breaks with \\
%\thanks{A footnote to the article title}%

\author{K. Bia{\l}as}
\affiliation{Institute of Physics, University of Silesia, 41-500 Chorz{\'o}w, Poland}
%\author{J. {\L}uczka}
%\affiliation{Institute of Physics, University of Silesia, 41-500 Chorz{\'o}w, Poland}
\author{J. Spiechowicz}
\email{jakub.spiechowicz@us.edu.pl}
\affiliation{Institute of Physics, University of Silesia, 41-500 Chorz{\'o}w, Poland}

\begin{abstract}
Active matter is one of the hottest topics in physics nowadays. As a prototype of living systems operating in viscous environments it has usually been modeled in terms of the overdamped dynamics. Recently, active matter in the underdamped regime has gained a place in the spotlight. In this work we unveil another remarkable face of active matter. In doing so we demonstrate and explain an inertia-induced mechanism of giant enhancement of transport driven by active fluctuations which does emerge neither in the overdamped nor in the underdamped limit but occurs exclusively in the strong damping regime. It may be relevant not only for living systems where fluctuations generated by the metabolism are active by default but also for artificial ones, in particular for designing ultrafast micro and nano-robots. Our findings open new avenues of research in a very vibrant field of active matter.
%attracted a quickly growing interest.
%become a hot topic
%attracted growing research interest.
%gained a place on the pedestal of interest
\end{abstract}

%abstrakt z 1 pracy o giant enhancement
%Active fluctuations are detected in a growing number of systems due to self-propulsion mechanisms or collisions with active environment. They drive the system far from equilibrium and can induce phenomena which at equilibrium states are forbidden by e.g. fluctuation-dissipation relations and detailed balance symmetry. Understanding their role in living matter is emerging as a challenge for physics. Here, we demonstrate a paradoxical effect in which a free particle transport induced by active fluctuations can be boosted by many orders of magnitude when the particle is additionally subjected to a periodic potential. In contrast, within the realm of only thermal fluctuations, the velocity of a free particle exposed to a bias is reduced when the periodic potential is switched on. The presented mechanism is significant for understanding nonequilibrium environments such as living cells, and it can explain from a fundamental point of view why spatially periodic structures known as microtubules are necessary to generate impressively effective intracellular transport. Our findings can be readily corroborated experimentally, e.g., in a setup comprising a colloidal particle in an optically generated periodic potential.

%\keywords{active fluctuations, Brownian particle, inertia, transport enhancement, periodic}%Use showkeys class option if keyword
                              %display desired
\maketitle

%\tableofcontents

\begin{quotation}
Complexity emerging at the interface of two or more distinct physical realms has always been a source of scientific challenges with fascinating results. It is enough to mention quantum mechanics as a bridge between the particle and wave face of matter \cite{bohm} or the still not fully solved problem of quantum to classical transition \cite{zurek}. One of the emanations of the latter is mesoscopic physics which constitutes a platform where both classical and quantum phenomena interact\cite{imry}. In this work we analyze in this context the active matter whose relevance for biological systems is emerging as a hot topic that is in focus of researchers across all branches of natural science, especially physics \cite{xu2024,trivedi2022,liu2021,prolife,topology,marchetti}. Specifically, we investigate an interface between its overdamped and underdamped dynamics and find a new instance of exotic transport feature.
\end{quotation}

\section{Introduction}
%In this work we analyze in this context the problem of active matter whose relevance for biological systems is emerging as a hot topic that is in focus of researchers across all branches of natural science, in particular physics \cite{xu2024,trivedi2022,liu2021,prolife,topology,marchetti}. 
Active matter such as e.g. self-propelled particles or micro- and nano-swimmers, is capable of harvesting energy from the environment into directed or persistent motion \cite{models,romanczuk,bechinger}. These systems are typically assisted by active nonequilibrium fluctuations which unlike thermal equilibrium ones are not limited by laws of physics like the fluctuation-dissipation theorem \cite{kubo,marconi} or detailed balance symmetry \cite{cates,gnesotto}. This fact opens new phenomenology of effects that remains unexplored to a large extent. For instance, biological motors such as kinesin or dynein exploit active fluctuations to enhance their directional motion along the microtubules \cite{ariga,ezber}.

The field of active matter was inspired by living systems whose understanding remains one of the greatest challenges in modern physics. %While operating in viscous environment and at inherently nonequilibrium conditions living organisms exhibit extraordinary properties like spontaneous motion or dynamical organization.
The latter operate under inherent nonequilibrium conditions in viscous environments \cite{bressloff,spiechowicz2023entropy}. For this reason active matter as a prototype of living system has usually been modeled in terms of the overdamped dynamics \cite{bechinger}. However, recently active systems in the underdamped regime, in which the damping is notably reduced so that the inertial effects play dominant role, 
%emerge as a hot topic in physics 
have gained a pole position at the pedestal of attention in physics \cite{hecht2024,hecht2022,lowen,antonov2024,caprini2024}.

In this work we unmask the another face of active matter, that so far has not been in the spotlight of research. In doing so we study the interface between the overdamped and underdamped active matter, namely the strong damping regime, in which dissipation dominates small but non-zero inertia. We demonstrate and explain an inertia-induced mechanism of giant enhancement of transport generated by active fluctuations which does not emerge in both overdamped and underdamped limits but occurs exclusively in the strong damping regime. 

Our results are relevant not only for living systems, e.g. molecular motors \cite{ariga,ezber}, which are exposed to both active and thermal fluctuations but also for designing ultrafast and efficient biologically inspired micro- and nano-robots \cite{bechinger}. The findings can open new avenues of research within a very vibrant field of active matter. It is supported by the fact that our results can be corroborated e.g. in recent experiments involving colloidal particles in optically generated periodic potentials \cite{park,paneru,paneru2023}.

This paper is organized as follows. In the following section we introduce the model of a Brownian particle driven by active fluctuations. Next, in Sec. III we discuss the inertia-induced mechanism for the giant enhancement of transport and explain its origin by constructing a relevant toy model. Sec. IV provides a summery and conclusions. In Appendix A we present the dimensionless formulation of the studied dynamics. In Appendix B and C we detail on the numerical simulation methodology as well as estimation of the transition probability from the system trajectory.

\section{Model}
We start our investigation with the overdamped dynamics of a free Brownian particle subjected to a static force $f$ formulated in terms of the following dimensionless Langevin equation
\begin{equation}
	\dot{x} = f + \sqrt{2D_T} \, \xi(t),
\end{equation}
where thermal fluctuations of intensity $D_T$ are modeled by Gaussian white noise with vanishing mean \mbox{$\langle \xi(t) \rangle = 0$} and the autocorrelation $\langle \xi(t)\xi(s) \rangle = \delta(t-s)$. The external force induces directed transport of the particle with the corresponding average velocity \mbox{$\langle v \rangle = \langle \dot{x} \rangle = v_0$}. Next, we put this forced system into a spatially periodic potential $U(x) = -\varepsilon \cos{x}$ so that it is described by the equation
\begin{equation}
	\dot{x} = -U'(x) + f + \sqrt{2D_T} \, \xi(t).
\end{equation}
Then it is expected that the motion of this passive forced particle is hampered by potential barriers with the height $2\varepsilon$ and consequently its averaged velocity is reduced in comparison to the free transport, i.e. $\langle v \rangle \le v_0$. We note that this observation holds true also when the full inertial dynamics is considered \cite{risken}. %However, it is not always the case. 

However, recently a paradoxical effect has been demonstrated in which a free-particle transport induced by active fluctuations $\eta(t)$ can be enormously boosted when the particle is additionally subjected to a periodic potential $U(x)$ \cite{pre}, i.e. $\langle v \rangle \gg v_0$. The dynamics of such a system is described by the Langevin equation
\begin{equation}
	\label{overdamped}
	\dot{x} = -U'(x) + \eta(t) + \sqrt{2D_T} \, \xi(t),
\end{equation} 
where for confrontation we additionally postulate $\langle \eta(t) \rangle = f$ so that for the free particle with \mbox{$U(x) = 0$} we have $\langle v \rangle = v_0$. These active nonequilibrium fluctuations may represent (i) a self-propulsion drive of an active particle harvesting energy from its environment \cite{models,romanczuk,bechinger,fodor2018} (ii) an active bath such as a suspension of active colloids acting on a passive system \cite{grober,lee2022,baule,baule2,eichhorn,bello2024,maggi2014} or (iii) an active particle immersed in an active bath itself. The above scenarios can be captured by fluctuations $\eta(t)$ in the form of a sequence of $\delta$-shaped pulses with random amplitudes $z_i$, i.e. as Poissonian white shot noise \cite{luczka,spiechowicz2014pre,bialas2020,mechanism}
\begin{equation}
    \eta(t)=\sum_{i=1}^{n(t)}z_i\delta(t-t_i),
\end{equation}
where $t_i$ are the arrival times of the Poissonian counting process $n(t)$ \cite{feller}, i.e. the probability for occurrence of $k$ impulses in the time interval $[0,t]$ is
\begin{equation}
    Pr\{n(t)=k\}=\frac{(\lambda t)^k}{k!}e^{-\lambda t}.
\end{equation}
Parameter $\lambda$ is the mean spiking rate determining how many $\delta$-pulses occur per unit time on average. Amplitudes $\{z_i\}$ are independent random variables drawn from a common probability distribution $\rho(z)$. Such model of active fluctuations $\eta(t)$ forms white noise with finite mean and covariance given by
\begin{subequations}
\begin{align}
\begin{split}
\langle\eta(t)\rangle&=\lambda\langle z_i\rangle,\\%=\lambda \zeta, \\
\end{split}\\
\begin{split}
\langle\eta(t)\eta(s)\rangle-\langle\eta(t)\rangle\langle\eta(s)\rangle&=2D_P\delta(t-s),\\%=\lambda(\sigma^2+\zeta^2)\delta(t-s), \\
\end{split}
\end{align}
\end{subequations}
where $D_P=\lambda \langle z_i^2\rangle/2$ is Poissonian white shot noise intensity. We assume that active fluctuations are uncorrelated with thermal ones, namely
\begin{equation}
\langle\eta(t)\xi(s)\rangle=\langle \eta(t) \rangle \langle \xi(s) \rangle=0.
\end{equation}

We note that the system given by Eq. (\ref{overdamped}) without active fluctuations $\eta(t)$ in the long time limit reaches an equilibrium state. It is important to distinguish this situation from the case in which the symmetry breaking bias of either deterministic or stochastic origin is applied to a system already in a nonequilibrium state. In such a case the average velocity of the particle in the periodic potential may be larger than the corresponding one for the free particle $\langle v \rangle > v_0$ even for a constant bias, see e.g. Ref. [\onlinecite{jung1991}]. 

\begin{figure}[t]
    \centering
    \includegraphics[width=0.95\linewidth]{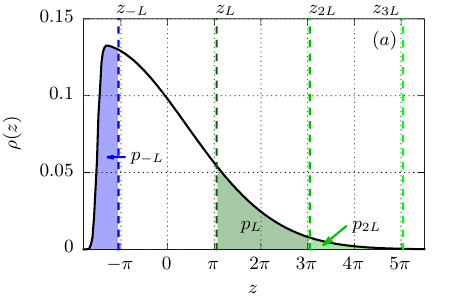}
    \includegraphics[width=0.95\linewidth]{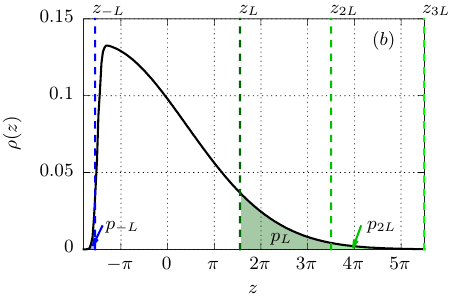}
    \caption{The probability density function $\rho(z)$ of active fluctuations $\eta(t)$ amplitudes $\{z_i\}$ with the critical amplitudes $z_{kL}$ and the corresponding probabilities $p_{kL}$ ($k \in \mathbb{N} \setminus \{0\}$) is presented for the particle mass $m=0.001$ and $m=0.017$ in panels $(a)$ and $(b)$, respectively. Other parameters are: the potential barrier $\varepsilon=100$, the mean amplitude $\zeta=1/20$, the variance $\sigma^2=13$ and the skewness $\chi=0.99$.}
    \label{fig1}
\end{figure}

In this study as a model of amplitude distribution $\rho(z)$ we pick the skew-normal density \cite{azz,rijal2022}. Such active fluctuations can represent e.g. the stochastic release of energy in a self-propelling mechanism or random collisions with suspension of active micro-swimmers forming an active bath. Skew-normal distribution is usually defined in terms of three parameters, namely a location $\mu$ representing the shift, a scale $\omega$ proportional to the variance, and a parameter $\alpha$ describing the shape. The probability density function $\rho(z)$ then reads
\begin{equation}
    \rho(z) = \frac{2}{\sqrt{2\pi \omega^2}}e^{-\frac{(z-\mu)^2}{2\omega^2}} \int_{-\infty}^{\alpha[(z-\mu)/\omega]} \frac{1}{\sqrt{2\pi}}e^{-\frac{s^2}{2}} ds.
\end{equation}
The quantities: $\mu$, $\omega$, and $\alpha$ are not very useful for applications. Fortunately they can be expressed in terms of the statistical moments of the distribution $\rho(z)$, i.e.,
its mean $\zeta = \langle z_i\rangle$, variance $\sigma^2 = \langle (z_i-\zeta)^2\rangle$ and a measure of asymmetry skewness $\chi = \langle (z_i-\zeta)^3\rangle/\sigma^3$ \cite{sp_generacja,sp_generacja2}
\begin{subequations}
\begin{align}
\begin{split}
\alpha&=\frac{\delta}{\sqrt{1-\delta^2}}, \\
\end{split}\\
\begin{split}
\omega&=\sqrt{\frac{\sigma^2}{1- 2\delta^2/\pi}}, \\
\end{split}\\
\begin{split}
\mu&=\zeta-\delta\sqrt{\frac{2\sigma^2}{\pi(1-2\delta^2/\pi)}},\\
\end{split}
\end{align}
\label{eq_S_def}
\end{subequations}
where $\delta$ is defined as
\begin{equation}
    \delta=\text{sgn}(\chi)\sqrt{\frac{|\chi|^{2/3}}{(2/\pi)\{[(4-\pi)/2]^{2/3}+|\chi|^{2/3}\}}}.
    \label{eq_S_delta}
\end{equation}
In Fig. \ref{fig1} we present the probability density function $\rho(z)$ of active fluctuations $\eta(t)$ amplitudes $\{z_i\}$. It is demonstrated for the mean amplitude $\zeta=1/20$, the variance $\sigma^2=13$ and the skewness $\chi=0.99$. 

%In addition, we have marked there the critical amplitudes $z_{kL}$ and the corresponding probabilities $p_{kL}$ for different mass $m$ of the particle and the potential barrier $\varepsilon = 100$. For $m=0.001$ (panel (a)) the critical amplitudes are close to the positions of potential minima as it is expected for the case of the overdamped limit. The reader can notice there a significant probability $p_{-L}$ for the particle being transported over the potential barrier into the negative direction. On the other hand, for the optimal mass $m_o=0.017$ the magnitude of critical amplitudes $|z_{kL}|$ increases so that the probability $p_{-L}$ becomes small.

The overdamped model given by Eq. (\ref{overdamped}) due to its simplicity suffers from several fundamental problems. The velocity exists therein only as an average observable. In such a model fundamental quantities like velocity fluctuations and its higher moments, kinetic energy, or thermodynamic efficiency cannot be defined \cite{jung,eichhorn2}. Not to mention the fact that it is always an approximation since in reality there are no massless systems. To address these serious problems we consider the full inertial dynamics which can be formulated in the dimensionless version as
\begin{equation}
	\label{inertial}
	 m\ddot{x} + \dot{x} = -U'(x) + \eta(t) + \sqrt{2 D_T} \, \xi(t),
\end{equation}
where $m$ represents the particle mass. The overdamped model (\ref{overdamped}) is recovered here by setting $m = 0$.

We shall investigate what is the impact of inertia $m$ on the average velocity $\langle v \rangle$ induced by active fluctuations $\eta(t)$ in the studied system. Since Eq. (\ref{inertial}) cannot be solved by analytical means we resorted to its precise numerical simulations \cite{spiechowicz2015cpc}. %It turns out that it is by no means minor as in the strong damping regime we discover entirely new mechanism of giant enhancement of transport driven by active noise that is present neither in the overdamped nor the underdamped dynamics. Our results are therefore relevant not only for artificial microscopic systems like a colloidal particle in an optically generated potential \todo{cite} but also biological ones such as e.g. living cells \todo{cite} where strongly damped kinetics is inherently affected by active fluctuations originating from metabolic activity. 
Below we limit ourselves to the following parameter regime: temperature of the system $D_T = 0.01$, the mean spiking rate $\lambda = 20$ of the active fluctuations $\eta(t)$ with the amplitudes $z_i$ sampled from the skew-normal distribution $\rho(z)$ of mean $\zeta = 1/20$, variance $\sigma^2 = 13$ and skewness $\chi = 0.99$.   

\section{Results}

\begin{figure}[t]
	\centering
	\includegraphics[width=0.95\linewidth]{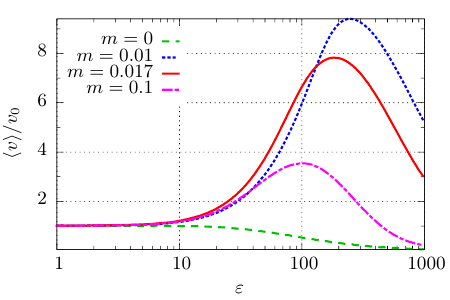}
	\caption{The rescaled average velocity $\langle v \rangle/v_0$ versus the periodic potential $U(x) = -\varepsilon \cos{x}$ barrier $\varepsilon$ is depicted for different mass $m$ of the particle.}
	\label{fig2}
\end{figure}
In Fig. \ref{fig2} we present the rescaled averaged velocity $\langle v \rangle/v_0$ versus the periodic potential $U(x)$ barrier $\varepsilon$ for different mass $m$ of the particle. The emergence of the giant enhancement of the free particle transport induced by active fluctuations $\eta(t)$ upon putting the particle in $U(x)$ is exemplified there. If the barrier vanishes $\varepsilon \to 0$ the average velocity of the particle tends to the value corresponding to free motion $\langle v \rangle \to v_0$. However, when the $\varepsilon$ grows we note the non-monotonic behavior of the rescaled velocity $\langle v \rangle/v_0$ versus $\varepsilon$ with the pronounced maximum observed for the optimal barrier $\varepsilon_o$. E.g. for $m = 0.01$ and $\varepsilon_o = 245$ the average velocity of the particle in the periodic potential is almost an order of magnitude larger than when it is free $\langle v \rangle/v_0 \approx 10$. Moreover, when the mass $m$ decreases both the optimal barrier $\varepsilon_o$ and the corresponding transport enhancement $\langle v \rangle/v_0(\varepsilon_o)$ grow. Remarkably, we also show that this effect does not occur for the overdamped system corresponding to $m = 0$. This fact suggests that its mechanism must be fundamentally different than the one detected in the overdamped limit.

\begin{figure}[t]
	\centering
	\includegraphics[width=0.95\linewidth]{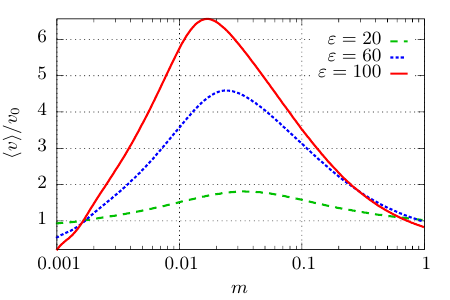}
	\caption{The rescaled average velocity $\langle v \rangle/v_0$ versus mass $m$ of the particle is shown for different barrier $\varepsilon$ of the periodic potential $U(x)$.}
	\label{fig3}
\end{figure}
The above findings are confirmed in Fig. \ref{fig3}, c.f. the limit $m \to 0$, where we show the same quantity of interest $\langle v \rangle/v_0$ but now as a function of mass $m$ of the particle for different barrier $\varepsilon$ of the periodic potential $U(x)$. Moreover, the observed effect of the giant enhancement of the free particle transport $\langle v \rangle/v_0 \gg 1$ %upon placing the particle in the periodic potential $U(x)$ is clearly induced by inertia as it disappears when mass vanishes $m \to 0$. Moreover, we can see that it 
is absent also for the underdamped regime $m \sim 1$ and for $m \gg 1$. In contrast, when the system is inertial but strongly damped $m \ll 1$ there exists a optimal mass $m_o$ for which the free transport enhancement $\langle v \rangle/v_0(m_o) > 1$ emerges. This fact unmasks a hidden face of active matter which as we demonstrate may exhibit exotic features that emerge exclusively in the strong damping regime. %It can open a new avenue of research within a very vibrant field of active matter.

To explain the mechanism of this phenomenon we first study a single trajectory of the particle in the periodic potential $U(x)$ with barrier $\varepsilon = 100$ for different mass $m$ \cite{sm2}. When the dynamics is overdamped $m = 0$ the particle waits for the arrival of the next $\delta$-pulse with the amplitude $z_i$ in the potential minimum %and therefore the system is highly localized. 
The reason for this behavior is its rapid relaxation towards the potential minimum. On the other hand, in the underdamped regime $m = 1$ %the system is delocalized, i.e. 
the particle oscillates around the minimum and it is kicked by $\delta$-pulse during the movement. An increase of mass $m$ visibly slows down the relaxation towards the minimum. %Due to this fact for the inertial dynamics a new transport mechanism emerges which we name inertial crossing. 
The particle can exploit the kinetic energy transferred from the last $\delta$-pulse to reduce %the energetic cost of the barrier crossing and when the next $\delta$-kick arrives the particle can
the effective potential barrier and when the next $\delta$-kick arrives overcome even multiple potential maxima. For the optimal mass $m_o = 0.017$ representing the strong damping regime we detect characteristics coming from both limits, i.e. %the system is almost localized as 
the particle tends to stay near the potential minimum and oscillates around it with a small amplitude. The relaxation is also much quicker than in the underdamped regime. %but the inertial crossing still emerges. 
Therefore for optimal mass $m_o = 0.017$ the system retains the advantages of both extreme regimes.

\subsection{Toy model}
To quantify these qualitative observations we construct a toy model of the underlying dynamics with the following approximations. (i) Since thermal fluctuations intensity $D_T$ is negligible in comparison to the potential barrier $\varepsilon$ we omit them. (ii) For vanishing mass $m \to 0$ the system is usually localized in the potential minimum $x = kL, k \in \mathbb{Z}$ (where $L = 2\pi$ is its spatial period) and consequently we assume that initially the particle rests $v(0) = 0$ therein $x(0) = 0$. (iii) We take into account only a single $\delta$-pulse of active fluctuations $\eta(t)$ with amplitude $z$ sampled from the skew-normal distribution $\rho(z)$. We now analyze what is the dependence of critical amplitude $z_c$, needed to transport the particle over the potential barrier $\varepsilon$ in the positive direction, on the mass $m$. Since the potential $U(x)$ is symmetric, the corresponding quantity in the opposite direction reads $z_{-c} = -z_c$.

For the overdamped dynamics described by Eq. (\ref{overdamped}) the first critical amplitude $z_c = z_L$ %needed to transport the particle across the neighboring potential barrier 
needed to overcome the potential barrier
is equal to the distance to the nearest maximum, i.e. $z_L = L/2 = \pi$. In this limit active fluctuations $\eta(t)$ act in the position space and the amplitude $z$ directly displace the particle by $\Delta x = z$. It is similar for the next amplitudes $z_{kL} = (2k-1)L/2 = (2k-1)\pi$, $k \in \mathbb{N} \setminus \{0\}$. Therefore there is a strict coupling between the critical values $z_{kL}$ and the spatial periodicity of the potential $U(x) = U(x + L)$. 

For the inertial dynamics given by Eq. (\ref{inertial}) the above toy problem can be formulated as
%\begin{equation}
%	\label{z_c}
	$m\ddot{x} + \dot{x} = -U'(x)$
%\end{equation}
with the initial particle position $x(0) = 0$ and velocity \mbox{$v(0) = z/m$}. Unfortunately, despite its simplicity, %Eq. (\ref{z_c}) 
this equation cannot be solved analytically and therefore we employed the numerical means to calculate the critical amplitudes $z_c$. In the Hamiltonian limit, i.e. without the dissipative term $\dot{x}$, all critical amplitudes $z_c = z_{kL}$ can be easily obtained from the conservation of energy $m v_c^2/2 = 2\varepsilon$ as $z_c = 2\sqrt{\varepsilon m}$. In such a case $z_c$ is coupled to the energetic barrier $\varepsilon$ rather than the spatial period $L$ of the potential $U(x)$. For the dissipative, inertial dynamics the critical amplitude $z_c$ must be greater due to the damping. %Its conservation reads $m v_c^2/2 = 2\varepsilon + E_d$ where $E_d$ stands for the energy dissipated by the latter contribution.

\begin{figure}[t]
	\centering
	\includegraphics[width=0.95\linewidth]{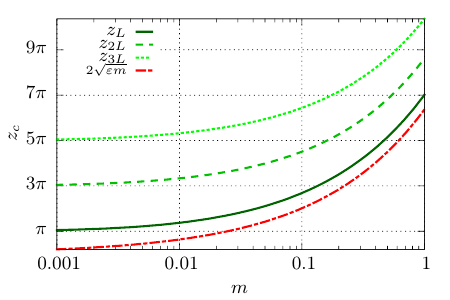}
	\caption{The critical amplitude $z_c$ of a single $\delta$-pulse of active fluctuations $\eta(t)$ needed to transport the particle over the first $z_L$, the second $z_{2L}$ and the third $z_{3L}$ potential $U(x)$ barrier as a function of mass $m$. The potential barrier reads $\varepsilon = 100$. The curve $2\sqrt{\varepsilon m}$ follows from the conservation of energy for the non-dissipative counterpart of the system.}
	\label{fig4}
\end{figure}
The result is shown in Fig. \ref{fig4} where we present the first $z_L$, the second $z_{2L}$ and the third $z_{3L}$ critical amplitude needed to transport the particle over the corresponding potential barrier as a function of mass $m$. For vanishing mass $m \to 0$ they tend to the values describing location of the potential maxima, as it is expected for the overdamped dynamics, see above. 
When the mass $m$ increases, the critical amplitudes $z_c$ also grows, but curiously the differences between them, e.g. $z_{2L} - z_L$ or $z_{3L} - z_{2L}$ decrease. It is so due to the increasing role of inertia which inhibits the dissipation of energy. The critical amplitudes $z_c = z_{kL}$, $k \in \mathbb{N} \setminus {0}$ allows us to calculate \emph{a priori} the partial probabilities $p_{\pm kL}$ for the particle being transported over the corresponding potential barrier in the positive and negative direction, namely
\begin{equation}
	 p_{\pm kL}= \pm \int_{\pm z_{kL}}^{\pm z_{(k+1)L}} \rho(z)dz, \quad k \in \mathbb{N} \setminus {0}.
\end{equation}
For instance, $p_L = \int_{z_L}^{z_{2L}} \rho(z) dz$ and $p_{-L} = \int_{-z_{2L}}^{-z_L} \rho(z) dz$. Note that e.g. $p_L$ is not equal to the total probability of crossing the nearest potential barrier in the positive direction which reads $p_L^\infty = \int_{z_L}^{\infty} \rho(z) dz$. 

In Fig. \ref{fig1} we present the critical amplitudes $z_{kL}$ and the corresponding probabilities $p_{kL}$ for different mass $m$ of the particle and the potential barrier $\varepsilon = 100$. For $m=0.001$ (panel (a)) the critical amplitudes are close to the positions of potential minima as it is expected for the case of the overdamped limit. The reader can notice there a significant probability $p_{-L}$ for the particle being transported over the potential barrier into the negative direction. On the other hand, for the optimal mass $m_o=0.017$ the magnitude of critical amplitudes $|z_{kL}|$ increases so that the probability $p_{-L}$ becomes small.
\begin{figure}[t]
	\centering
	\includegraphics[width=1.0\linewidth]{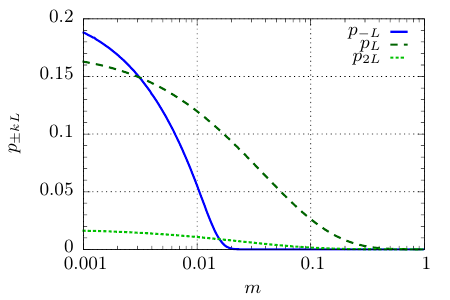}
	\caption{The dependence of the partial probabilities $p_{\pm kL}$, \mbox{$k \in \mathbb{N} \setminus {0}$}, for the particle being transported over the corresponding potential $U(x)$ barrier with height $\varepsilon = 100$ on the particle mass $m$.}
	\label{fig5}
\end{figure}

The dependence of $p_{\pm kL}$ on the particle mass $m$ for the potential barrier $\varepsilon = 100$ is depicted in Fig. \ref{fig5}. When the mass $m$ is increased the probabilities $p_{\pm kL}$ for $k > 1$ quickly decay since then the critical amplitudes $z_{\pm kL} = \pm z_{kL}$ grows as well. For this reason we show only the lowest order ones. In the overdamped limit \mbox{$m \to 0$} we note that the difference $(p_L + p_{2L}) - p_{-L}$ between the probabilities corresponding to processes transporting the particle over the potential barrier into the positive and negative direction is very small so that the average velocity $\langle v \rangle$ of the particle vanishes as well, c.f. Fig. \ref{fig3}. The same observation holds true in the underdamped regime $m \to \infty$ when $\delta$-pulses of active fluctuations $\eta(t)$ do not provide enough energy to overcome the potential barrier. However, for the strong damping case $m \ll 1$ the probability $p_L$ is significantly larger than $p_{-L}$ so that the enhancement of transport in the positive direction may emerge.

\begin{figure}[t]
    \centering
    \includegraphics[width=0.95\linewidth]{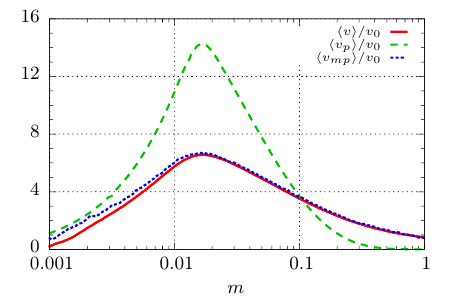}
    %\caption{Rescaled average velocities calculated with different methods vs. mass $m$, for barrier height $\varepsilon=100$, mean spiking rate $\lambda=20$, average impulse amplitude $\zeta=1/20$, variance $\sigma^2=13$ and skewness $\chi=0.99$. Thermal noise intensity was set to $D_T=0.01$ for $v_{sim}/v_0$ and $\langle v\rangle/v_0$.}
    \caption{Comparison of the rescaled average velocity of the particle as a function of its mass $m$ obtained with different approaches: $\langle v \rangle/v_0$ numerical simulations of the Langevin \mbox{Eq. (\ref{inertial})}; $\langle v_p \rangle/v_0$ the phenomenological toy model Eq. (\ref{toy}) and $\langle v_{mp} \rangle/v_0$ the modified toy model, see the text in the paragraph before the last one. The potential barrier reads \mbox{$\varepsilon = 100$}.}    
    \label{fig6}
\end{figure}
The outcomes of the above toy model can be confrontated with the results of precise numerical simulations shown in Fig. \ref{fig3}. For this purpose we express the average velocity of the particle as
\begin{equation}
	\label{toy}
	\langle v_p \rangle  = v_\lambda \sum_{k = 1}^{\infty} k p_k = v_\lambda \sum_{k = 1}^{\infty} k \left( p_{kL} - p_{-kL} \right),
\end{equation}
where $v_\lambda = L/(1/\lambda) = L\lambda = 2\pi \lambda$ is the characteristic velocity related with the considered dynamics, $L$ is the spatial period of the potential and $\lambda$ is the mean spiking rate of active fluctuations $\eta(t)$. %, i.e. the average number of $\delta$-pulses per unit time. 
The comparison of $\langle v_p \rangle$ with $\langle v \rangle$ is shown as a function of mass $m$ in \mbox{Fig. \ref{fig6}}. While there are quantitative differences between these two curves their qualitative behavior is the same. Most importantly, our phenomenological toy model captures the emergence of the giant enhancement of free-particle transport driven by active fluctuations $\eta(t)$ when it is exposed to the periodic potential $U(x)$. Moreover, remarkably, it correctly predicts the optimal mass $m_o = 0.017$ corresponding to the strong damping regime for which the amplification $\langle v \rangle/v_0$ is most pronounced. %It means that the mechanism standing behind this paradoxical behavior lies in the particular influence of inertia $m$ on probabilities $p_{\pm kL}$ for the particle being transported over the potential $U(x)$ barrier. It is realized by modification of the critical amplitudes $z_c$ of active fluctuations $\eta(t)$ needed to trigger such a process. 

The origin of discrepancy between the curves $\langle v \rangle/v_0$ and $\langle v_p \rangle/v_0$ presented in Fig. \ref{fig6} lies in the employed approximations of the constructed toy model. We considered only a single $\delta$-pulse of active fluctuations $\eta(t)$ and assume that initially the particle rests at the potential minimum. These conditions are obeyed only for vanishing mass $m \to 0$ and rare $\delta$-pulses $\lambda \to 0$. In the studied regime $\lambda = 20$ so the latter is not fulfilled, however, the agreement between $\langle v \rangle/v_0$ and $\langle v_p \rangle/v_0$ is nevertheless quite good for $m \to 0$. In practice, when a $\delta$-pulse arrives the inertial particle is almost always in motion around the potential minimum which can lower the effective potential barrier. It implies that the already accumulated momentum allows to overcome the potential even though it would not be possible if the initial position of the particle would correspond to the minimum. This fact explains why for large mass $m \sim 1$ the average velocity obtained from the toy model $\langle v_p \rangle/v_0$ is significantly smaller than $\langle v \rangle/v_0$. The above limitations of the presented phenomenological model can be surmounted when the expression for the average velocity given in Eq. (\ref{toy}) is considered as %$\langle v_{mp} \rangle = v_\lambda (p_+ - p_-)$ where $p_+$ and $p_-$ are
\begin{equation}
	\label{mod_toy}
	\langle v_{mp} \rangle = v_\lambda (p_+ - p_-),
\end{equation}
%where $p_\pm = n_\pm/(\lambda \mathsf{T})$ are the probabilities for the potential barrier crossing events in the positive and negative direction, respectively, 
where $p_+$ and $p_-$ are calculated directly from the system trajectory (for details see Appendix C). %For this purpose one needs to track the number $n_\pm$ of these incidents during the simulation which lasts for $\mathsf{T}$ dimensionless units of time. 
In Fig. \ref{fig6} one can observe very good agreement between $\langle v \rangle/v_0$ and $\langle v_{mp} \rangle/v_0$ which proves that the employed approach is essentially correct.

\section{Conclusions}
In conclusion, we discovered a new mechanism of inertia-induced giant enhancement of free transport driven by active fluctuations upon placing the system in a periodic potential. Its origin lies in the impact of inertia on the probability for the system to overcome the potential barrier by alteration of the critical amplitudes of active fluctuations needed to trigger such a process. Most remarkably, we show that this effect does not occur neither in the overdamped nor the underdamped dynamics but only in the strong damping regime. This fact unmasks the remarkable face of active matter which as we demonstrate may exhibit exotic transport features that emerge exclusively in the strong damping regime that so far has not been in the spotlight of research. Therefore we expect that it will open new avenues of study within a very vibrant field of active matter.

%In summary, we have shown how adding inertia to the system affects the giant transport enhancement of Brownian particle in periodic potential exposed to active fluctuations. Not only transport in the inertial system can be more efficient, but also using the overdamped limit can lead to drastically different predictions in extreme regimes, e.g., for considerable barrier heights $\varepsilon$. The addition of inertia changes the coupling of active fluctuations to spatial period to energy coupling. We introduced a simplified approximate expression for average velocity based on transition probabilities towards potential minima. Although it does not provide an accurate estimate of the magnitude of average velocity $\langle v\rangle<v_P$, it predicts what mass $m$ is optimal for given active fluctuations $\eta(t)$ and periodic potential $U(x)$. Our findings can be corroborated experimentally in both biological systems \cite{ezber,ariga} immersed \textit{in situ} in a sea of thermal and active fluctuations or artificial microscopic setups, e.g., Josephson junctions \cite{spiechowicz2015chaos} or colloidal particles in optical potentials\cite{park,paneru,paneru2023}.
\section*{Acknowledgement}
This work has been supported by Grant NCN No. 2022/45/B/ST3/02619 (J.S.)

\section*{Conflict of interest}
The authors have no conflicts to disclose.

\section*{Data availability statement}
The data that support the findings of this study are available from the corresponding author upon reasonable request.

\appendix

\section{Dimensionless Langevin dynamics}
Transformation of the equation describing the considered model into its corresponding dimensionless form allows to reduce the number of system parameters as well as makes the theoretical predictions independent of a specific experimental setup so that they can be applied to dynamics of any system sharing the same dimensionless representation. The latter fact is particularly important for the experimental corroboration of the results since one can pick a setup which is particularly convenient for such purpose and overcome many technical difficulties that can arise for other ones.

We start with the Langevin equation for an inertial Brownian particle in a periodic potential $U(x)$ driven by active fluctuations $\eta(t)$

%\begin{equation}
%	\Gamma \dot{x} = - U'(x) + \eta(t) + \sqrt{2 \Gamma k_B T}\, \xi(t)
 %   \label{seq_1}
%\end{equation}
%and Langevin equation for the corresponding inertial system
\begin{equation}
    M\ddot{x} + \Gamma \dot{x} = -U'(x) + \eta(t) + \sqrt{2 \Gamma k_B T}\, \xi(t)
    \label{seq_2}
\end{equation}
Here $M$ represents the particle mass, $\Gamma$ is the friction coefficient, $k_B$ the Boltzmann constant and $T$ denotes thermostat temperature. The spatially periodic potential $U(x)$ is assumed to be in the form
\begin{equation}
	U(x) = E \sin{\left( 2 \pi \frac{x}{L} \right)},
\end{equation}
where $E$ is a half of the potential barrier height and $L$ stands for its period. Thermal fluctuations $\xi(t)$ are modeled as $\delta$-correlated Gaussian noise of vanishing mean, i.e.
\begin{equation}
	\langle \xi(t) \rangle = 0 \quad \langle \xi(t)\xi(s) \rangle = \delta(t-s).
\end{equation}
\\

We rescale the particle position and time in the following way
\begin{equation}
\begin{split}
	 \hat{x} &= 2\pi \frac{x}{L}, \quad \hat{t}=\frac{t}{\tau_0}, \\
  \tau_0 &= \frac{L}{100 v_{D_0}} = \frac{L}{100 D_0/L} = \frac{L^2}{100 k_B T/\Gamma}
  \end{split}
\end{equation}
where $v_{D_0}$ is the characteristic velocity corresponding to free thermal diffusion $D_0 = k_B T/\Gamma$. The additional factor $100$ is introduced in the denominator due to technical reasons outlined below, see Appendix B. Under such a choice of scales, Eq. (\ref{seq_2}) becomes
%\begin{equation}
%	\dot{\hat{x}} =-\hat{U}'(\hat{x}) + \hat{\eta}(\hat{t}) + \sqrt{2D_T}\,\hat{\xi}(\hat{t})
%	\label{dimless_model}
%\end{equation}
%and Eq. (\ref{seq_2})
\begin{equation}
	m\ddot{\hat{x}}+\dot{\hat{x}} =-\hat{U}'(\hat{x}) + \hat{\eta}(\hat{t}) + \sqrt{2D_T}\,\hat{\xi}(\hat{t}).
	\label{dimless_model_inert}
\end{equation}
In this scaling, dimensionless mass $m$ reads
\begin{equation}
    m=\frac{\tau_1}{\tau_0}=\frac{100M k_B T}{\Gamma^2 L^2},
\end{equation}
where the second characteristic time is $\tau_1=M/\Gamma$. The rescaled potential
\begin{equation}
	\hat{U}(\hat{x}) = \frac{1}{100 k_B T} U\left( \frac{L}{2\pi} \hat{x} \right) = \varepsilon \sin{\hat{x}}
\end{equation}
possesses the spatial period $2\pi$ and the variable barrier height $\varepsilon = E/(100 k_B T)$. The above scaling procedure allows us to reduce the number of free parameters by
\begin{equation}
	\gamma = 1, \quad D_T = 0.01.
\end{equation}
The dimensionless thermal noise reads 
\begin{equation}
	\hat{\xi}(\hat{t}) = \frac{L}{2\pi} \frac{1}{100 k_B T}\, \xi(\tau_0 \hat{t}).
\end{equation}
It exhibits the same statistical properties as the dimensional one, i.e., it has the same vanishing mean and correlation function. The rescaled active fluctuations $\eta(t)$ are
\begin{equation}
	\hat{\eta}(\hat{t}) = \frac{L}{2\pi} \frac{1}{100 k_B T}\, \eta(\tau_0 \hat{t}).
\end{equation}
As in the main text, only the dimensionless quantities are used. We omit there the hat notation.

% The \nocite command causes all entries in a bibliography to be printed out
% whether or not they are actually referenced in the text. This is appropriate
% for the sample file to show the different styles of references, but authors
% most likely will not want to use it.
%\nocite{*}

\section{Numerical simulation}
The analytical solutions of Eq. (\ref{dimless_model_inert}) nor the corresponding Fokker-Planck-Kolmogorov-Feller one is not attainable. For this reason we resorted to precise numerical simulations. $2^{16}$ system trajectories were simulated using the modified Euler-Maruyama method \cite{kim2007,Platen} that allows to take into account active fluctuations $\eta(t)$. The simulations last up to time $t=1000$ for which the asymptotic state was reached and then the ensemble average was performed to obtain the quantities of interest. Such a task is ideal for parallel computation, therefore we used graphical processing units (GPUs) for calculations. It allows us to speedup them by a factor of almost ten thousand. For more details on the Monte Carlo integration scheme and the speedup on GPUs we refer the readers to \cite{spiechowicz2015cpc}. Moreover, we note that the growth of the dimensionless mean spiking rate $\hat{\lambda} = \tau_0 \lambda$ of Poissonian white shot noise $\hat{\eta}(\hat{t})$ not only degrades the speed of calculations, but also escalates the errors of the employed numerical integration scheme \cite{kim2007}. In order to reduce these drawbacks, factor $100$ was introduced in the definition of characteristic time scale $\tau_0$, which allowed us to significantly limit the spiking frequency $\hat{\lambda}$ range needed in this study. 

\begin{figure}
    \centering
    \includegraphics[width=\linewidth]{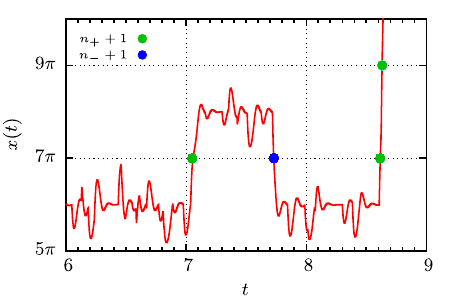}
    %\caption{Exemplary transition counting scheme for single trajectory for mass $m=0.017$, barrier height $\varepsilon=100$,  mean spiking rate $\lambda=20$, mean amplitude $\zeta=1/20$, variance $\sigma^2=13$ and skewness $\chi=0.99$. Colored dots correspond to the increase in counters $n_{\pm}$.}
    \caption{A single exemplary trajectory of the system for the mass $m=0.017$, the barrier height $\varepsilon=100$, the mean spiking rate $\lambda=20$, the average amplitude $\zeta=1/20$, the variance $\sigma^2=13$ and skewness $\chi=0.99$. Colored dots correspond to the events of the periodic potential barrier crossing which translates to the increase in the counter $n_{\pm}$.}
    \label{fig:S2}
\end{figure}

%\section{Method of obtaining transition probabilities from Monte Carlo simulations}
\section{Estimation of the transition probabilities $p_\pm$ from the system trajectory}
The probabilities $p_\pm$ for the particle to cross the potential $U(x)$ barrier either in the positive or negative direction as a result of $\delta$-spike of active fluctuations $\eta(t)$ can be calculated from a single system trajectory. For this purpose one needs to track the particle position $x(t)$ modulo the spatial period $L$ of the potential $U(x)$. For our case $U(x) = -\varepsilon \cos{x}$ it can be done via the following floor function
%In the main paper, besides transition probabilities $p_{\pm kL}$ in a simplified phenomenological model,
%\begin{equation}
%    p_{\pm kL}=\pm\int^{\pm z_{(k+1)L}}_{\pm z_{kL}}\rho(z)dz,\quad k\in \mathbb{N} \backslash 0
%\end{equation}
%we also presented analogous probabilities $p_{\pm}$ for the more complex system. Here, we show in more detail how to obtain them. We start by simulating a trajectory of a single particle. Simulation time $t$ must not be too long to reduce numerical errors from division by $\pi$. Instead, we suggest repeating the simulation for more trajectories for the same system parameters to increase the sample size of barrier crossing events. During the numerical simulation we check in which basin of attraction is the particle at each simulation step. To do that, we use the floor function $\lfloor \cdot \rfloor$ to assign an integer $i$ to particle position $x$, determining between which maxima it is, for our potential $U(x)=-\varepsilon \cos(x)$ it has the following form:
\begin{equation}
    l =\left \lfloor{\frac{x+\pi}{2\pi}}\right \rfloor.
\end{equation}
During the simulation of system trajectory one needs to increase the number $n_{+}$ of positive and $n_-$ negative transitions each time the integer $l$ increases or decreases, respectively. We regain transition probabilities $p_{\pm}$ by rescaling the counters $n_{\pm}$ by the product $n = \lambda \mathsf{T}$ of the mean spiking rate $\lambda$ and the time span $\mathsf{T}$ of the trajectory, i.e.
\begin{equation}
    p_{\pm}=\frac{n_{\pm}}{n}.
\end{equation}
In Fig. \ref{fig:S2} we present this procedure for a short part of the system trajectory. The particle crossing events are marked there by the corresponding colored dots.

\end{document}